# Brain Inspired Computing Approach for the Optimization of the Thin Film Thickness of Polystyrene on the Glass Substrates

Akshansh Mishra[a], Devarrishi Dixit [b]

[a] Centre for Artificial Intelligent Manufacturing Systems, Stir Research Technologies, India

[b]Department of Materials Science Engineering, Christian Albrechts University zu Kiel, Germany

**Abstract:** Advent in machine learning is leaving deep impact on various sectors including material science domain. The present paper highlights the application of various supervised machine learning regression algorithms such as polynomial regression, decision tree regression algorithm, random forest algorithm, support vector regression algorithm and artificial neural network algorithm to determine the thin film thickness of Polystyrene on the glass substrates. The results showed that polynomial regression machine learning algorithm outperforms all other machine learning models by yielding the coefficient of determination of 0.96 approximately and mean square error of 0.04 respectively.

## 1. Introduction

Spin coating is a technique for depositing thin films on flat substrates. It is highly relevant for depositing films on silicon wafers for applications in semiconductor technology and photolithography. It is based on using the centrifugal force exerted by a rotational movement of the substrate on which a material solution is placed which upon evaporation of the solvent creates a thin film [1-4]. There are several requirements for the application of the spin coating. The thin-film material needs to be dissolved in a volatile solvent. The substrate material needs to be wettable by the solution and sufficiently flat. The technique of static spin coating is depicted schematically in Figure 1. The process shown in Figure is divided in to four steps i.e. firstly it represents deposition process, secondly it represents spin-up process, thirdly it represents Spin-off process and last process is evaporation. First, a droplet of the material solution is placed on the substrate. Then the substrate gets accelerated in order to achieve the desired rotational frequency. During this process, a significant amount of solution is flung off the substrate, which thins the deposited droplet. When the acceleration phase is finished and the final rational frequency is reached, this frequency is held for a certain amount of time to exert a constant centrifugal force on the solution [5-7]. Because of the centrifugal force, the solution can spread evenly on the surface which leads to the desired uniform film thickness. Still remaining solvent evaporates and at the end of the process, a thin film on the substrate is deposited. In general the thinning behaviour is determined by various aspects:

1.  spinning time

2. rotational frequency

3. viscosity of the solution

4. concentration of the material in the solution

5. evaporation rate

6. wetability of the substrate by the solution

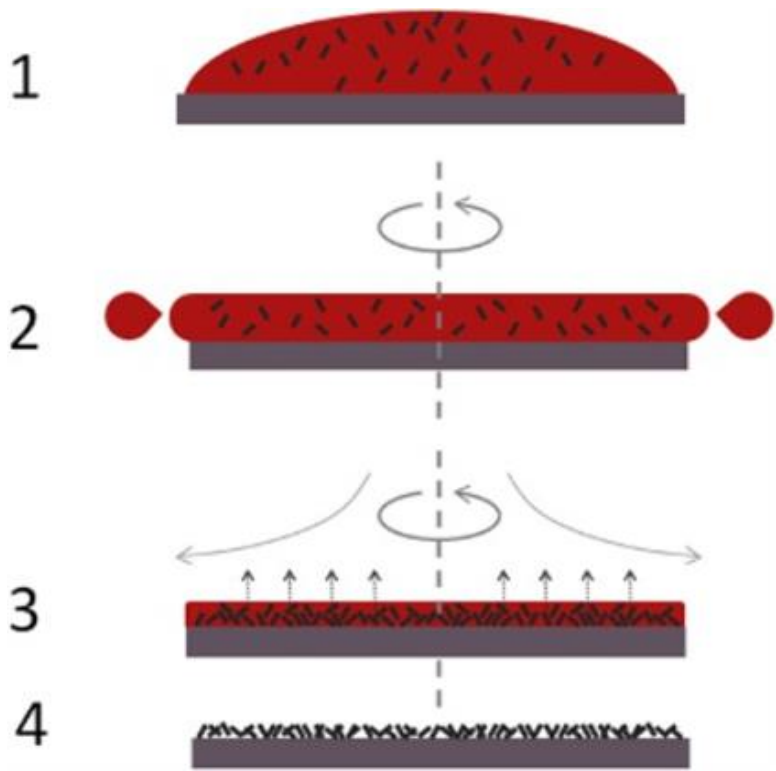

Figure 1. Schematically illustration of the steps involved in spin coating process [8]: (1) deposition of droplet; (2) acceleration to final rational frequency, (3) final rotational frequency and evaporation of solvent; (4) rotation stop

The spinning time determines how long the centrifugal forces are exerted on the solution and for how long the solvent can evaporate. The rotational frequency determines the centrifugal force acting on the solution therefore it, too, influences the resulting thin film thickness. The viscosity of the solution determines its flow properties. Therefore it has a significant influence on the resulting thickness uniformity. It can be stated, the more material is present in the solution the higher the viscosity. Therefore the concentration of the material has an effect on thinning behavior, too. Also, the evaporation rate determines the thickness of the resulting film, because when the solvent is evaporating the solution gets richer in solute which increases the viscosity of the fluid. Furthermore, the wettability of the substrate with the material solution plays an important role in the deposition of the thin film. Like already mentioned the wetting of the substrate has to be guaranteed by the solution. Therefore the interface energy between solution and substrate has to be chosen in such a way that wetting becomes possible. Nowadays, machine learning has become more dominant in manufacturing and material science domain. Machine learning contributes by reducing the cost and time of the experiment and results higher accuracy. Wakabayashi et al.[9] used Bayesian optimization-based machine learning algorithm for thin-film growth. The results showed that the implementation of the machine learning algorithm reduced experiment time and also it reduced the cost of the experiment. Ding et al. [10] used an artificial neural network-based machine-learning algorithm to optimize the atomic layer deposition cycle time. It was concluded that the developed algorithm can be used for enhancing the various industrial manufacturing processes. Greco et al. [11] determined the density, thickness, and roughness of different organic compounds such as α-sexithiophene, di-indenoperylene, and copper(II) phthalocyanine by using a simple artificial neural network. The machine learning model yields good results with a mean absolute percentage error of 8–18%. Banko et al. [12] used generative machine learning to predict structure zone diagrams for thin films synthesis. The results showed that the generative machine learning can be used for the optimization of process parameters and chemical composition to obtain a desired microstructure. A NIMS-University of Tokyo research group has developed a machine learning technique which can be used to expedite the process of determining optimum conditions for the fabrication of high-quality thin films by reducing the number of material samples needed to be evaluated by up to 90% compared to currently available thin film fabrication methods. The technique may be used to reduce the cost of developing many different types of thin film materials [13]. The schematic representation of the experiment carried out is shown in the Figure 2.

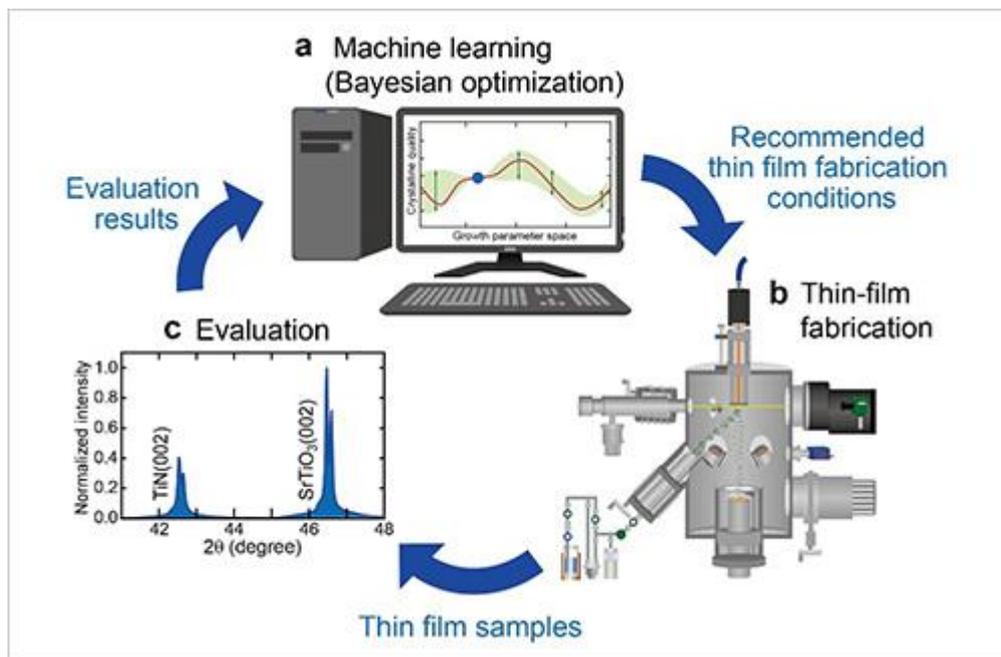

Figure 2. Schematic illustration of a machine-learning-integrated closed-loop process to optimize thin film fabrication parameters [13]

Panfilova et al. [14] modeled the metal islands thin films growth while vacuum evaporation process by using Artificial Neural Network algorithm as shown in the Figure 3.

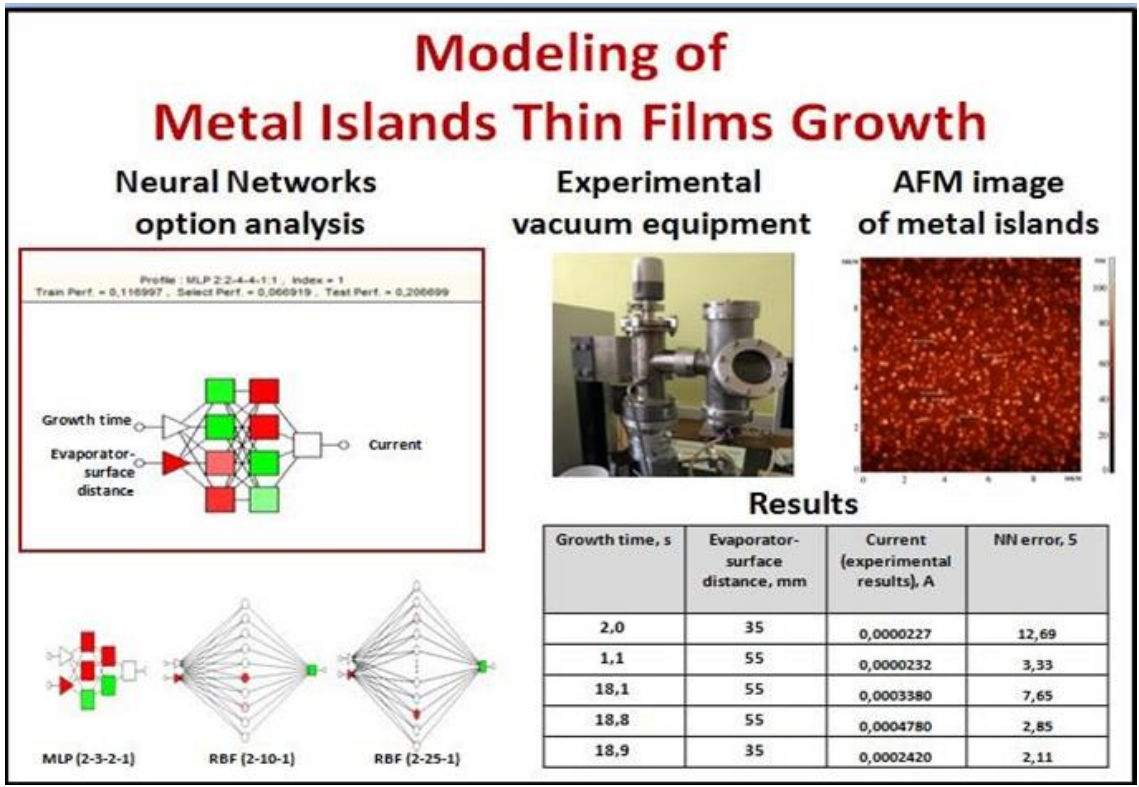

Figure 3. Modeling of the metal islands thin films growth while vacuum evaporation process [14].

From the previous research work, it is observed that there are few papers which implemented Machine Learning algorithms in Thin films fabrication. In the present work, supervised machine learning algorithms are developed by using Python programming for predicting the thin film thickness of Polystyrene on the glass substrates.

## 2. Experimental Procedure

In order to analyze the film thickness dependency on the angular velocity, spin coating is carried out at four different angular frequencies (in revolutions per minute, rpm): 500 rpm, 1000 rpm, 1500 rpm, 2000 rpm, and 3000 rpm. The effect of the polymer concentration on the film thickness is assessed by using two different solutions, which contain concentrations of 2 %, 3 %, 4%, and 8% of polystyrene dissolved in toluene. In order to evaluate the reproducibility, several samples are spin-coated with the 2 %-solution at a frequency of 1000 rpm. As a substrate glass-wafers with a size of 20 mm x 20 mm are used.

For the experiment, the spin coater Model WS-650MZ-23NPPB is used. In order to create a reproducible atmosphere during the spinning nitrogen gas is pumped into the spin coater. The glass-wafers are placed on the chuck and then fixed via a vacuum. After dropping 600 µl of the solution on the substrate with an Eppendorf pipette the spin process is started. The process consists of two different steps. First, the substrate is accelerated for 15 s to the adjusted velocity. In the second step, the speed is held constant for 60 s. The film thickness is measured by the profilometer DektakXT of Bruker. Therefore every sample is scratched carefully with tweezers in order to remove a small part of the film without damaging the substrate. After that, the samples are placed in the profilometer and the film thickness is measured at three different positions. To this end, a diamond tip is scanning over the scratch. By measuring the height difference of the scratched and the unscratched section the film thickness is determined. The experimental dataset is shown in Table 1.

The Python libraries which are imported for constructing and executing the Machine Learning algorithms were Numpy, Matplotlib, Seaborn, Pandas, Tensorflow, and Keras. Figure 4 shows the hierarchy of the experimental procedure subjected to the CSV dataset. In our present study, supervised machine learning algorithms such as Polynomial Regression, Support Vector Regression, Decision Tree Regression, Random Forest Regression and Deep Artificial Neural Network were implemented for optimizing the thickness of the thin films.

Table 1. Experimental Dataset

| Concentration | Angular Speed (rpm) | Thickness (nanometer) |
|---|---|---|
| 2 | 1000 | 751.00 |
| 2 | 1500 | 644.00 |
| 2 | 2000 | 520.00 |
| 2 | 3000 | 443.00 |
| 2 | 3000 | 437.00 |
| 2 | 3000 | 452.00 |
| 3 | 1000 | 852.00 |
| 3 | 1500 | 592.00 |
| 3 | 2000 | 527.00 |
| 3 | 3000 | 526.00 |
| 8 | 500 | 5898.70 |
| 8 | 1000 | 4811.70 |
| 8 | 1500 | 4850.00 |
| 8 | 2000 | 4300.30 |
| 4 | 500 | 668.30 |
| 4 | 1000 | 702.70 |
| 4 | 1500 | 587.70 |
| 4 | 2000 | 733.30 |

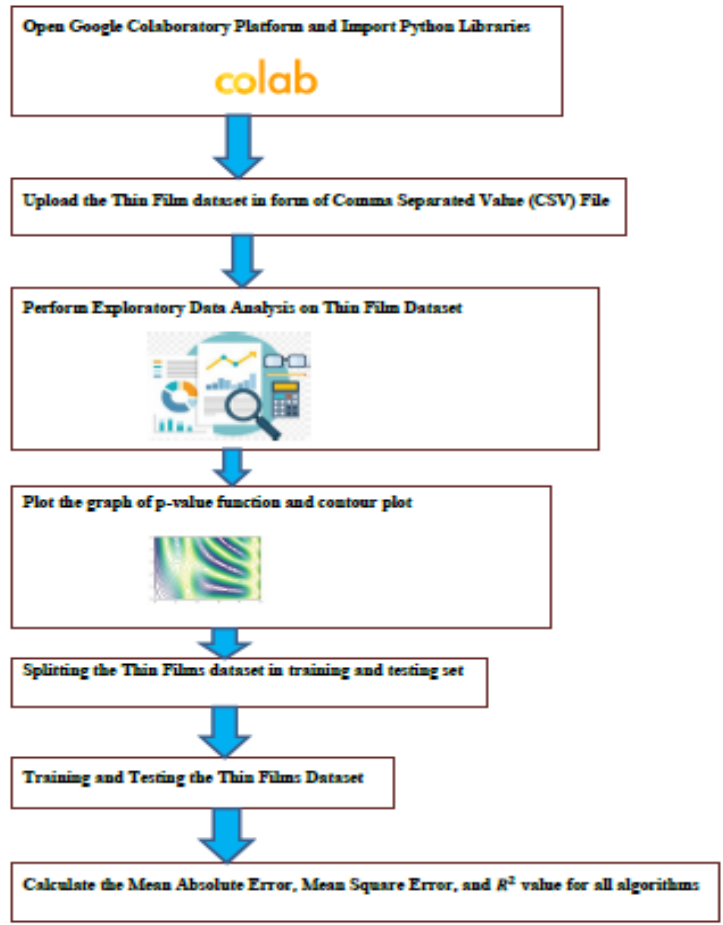

Figure 4. Process flow chart of the implementation of machine learning algorithms

## 3. Results and Discussion

### 3.1 Exploratory Data Analysis

Firstly, we have explored the relation which the features shared with the target variable. Accordingly, the features were dropped which have no relation with the target variable. From Table 2 we can see the data's distribution and judge whether we need to normalize our data or not. We also get other statistics using the table.

Table 2. Statistical model of the experimental dataset

|       | Concentration | Angular Speed (rpm) | Thickness (nanometer) |
|-------|---------------|---------------------|-----------------------|
| count | 18.00000      | 18.000000           | 18.000000             |
| mean  | 4.00000       | 1722.222222         | 1572.038889           |
| std   | 2.32632       | 844.048755          | 1890.746043           |
| min   | 2.00000       | 500.000000          | 437.000000            |
| 25%   | 2.00000       | 1000.000000         | 526.250000            |
| 50%   | 3.00000       | 1500.000000         | 656.150000            |
| 75%   | 4.00000       | 2000.000000         | 826.750000            |
| max   | 8.00000       | 3000.000000         | 5898.700000           |

**3.2 Checking Null Values in the dataset**

Secondly, the check_null( ) function is used to check the number of null values in the dataset. The null values are replaced by mean.

**3.3 Plotting Graph of p-Value Function and Contour Plot**

The plot_graph_pvalue ( ) function plots a line plot between given variables and prints the p-value and Pearson values. The contour_plot ( ) function plots a contour plot for the given variables.

Figure 5 shows the plot between the concentration and thickness value. The obtained p-value and Pearson value for the given parameters is 0.0000 and 0.942 respectively. From the p-value and Pearson value, we can clearly interpret that the value of concentration is highly correlated with the film thickness. From the graph, we can see the Pearson's predictions come to life as we see the film thickness values start to sudden increase with increasing concentration after the value of 4. Figure 6 shows the contour plot of shoulder diameter and UTS.

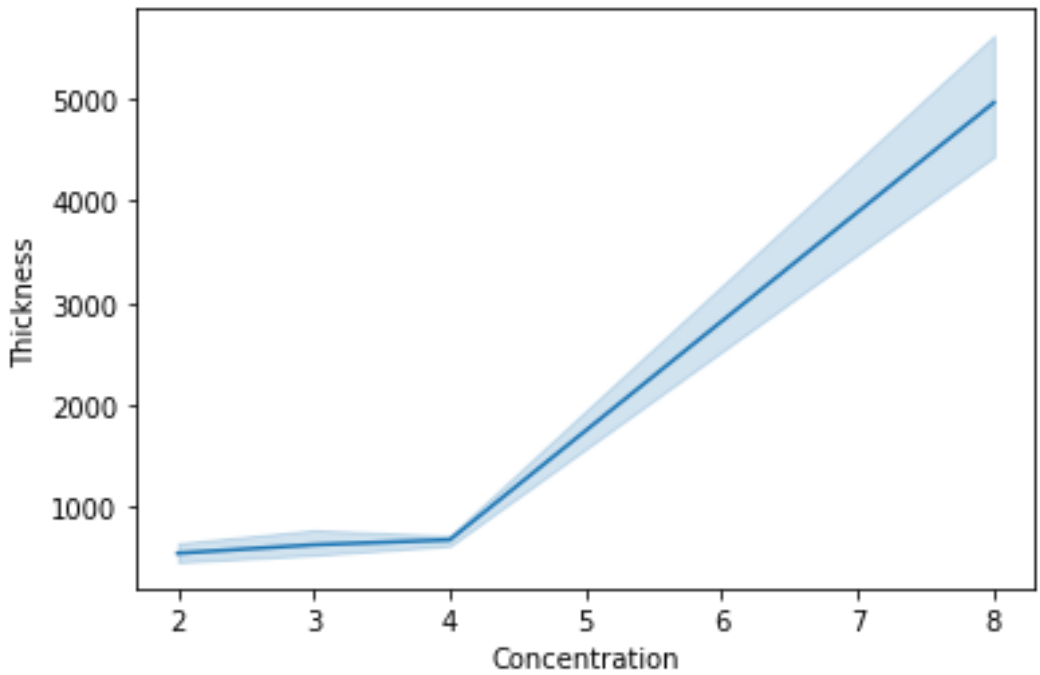

Figure 5. Relationship between the concentration value and the thickness value

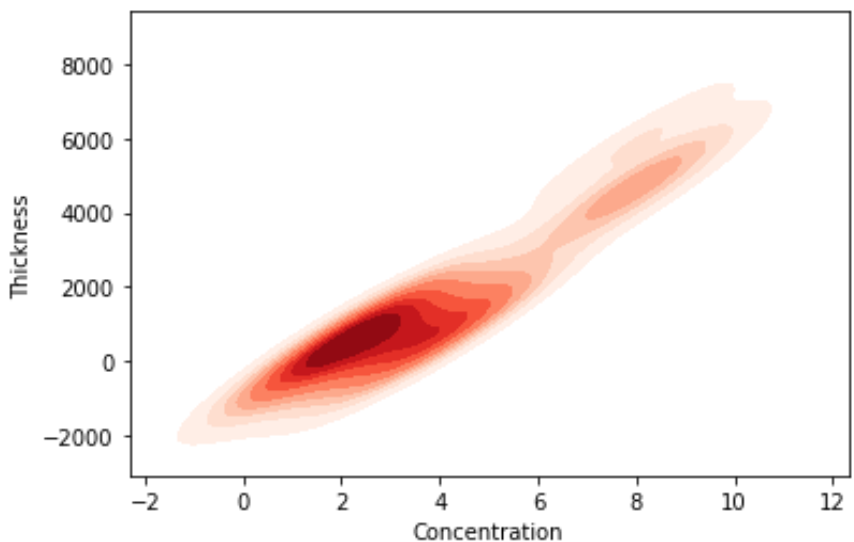

Figure 6. Contour plot between the thickness value and concentration value

Figure 7 shows the plot between the angular speed and thickness value. The obtained p-value and Pearson value for the given parameters is 0.10940 and - 0.390 respectively. It is observed that the film thickness value starts decreasing with increase in angular speed value. Figure 8 shows the contour plot between the angular speed value and film thickness value.

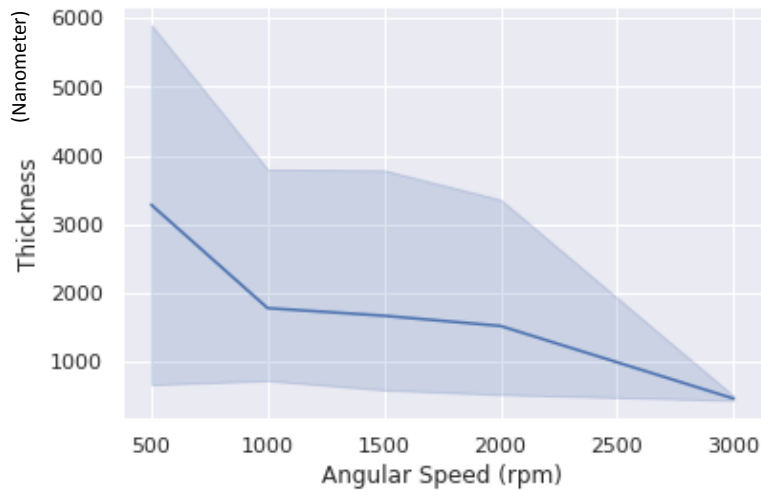

Figure 7. Relationship between the film thickness and angular speed.

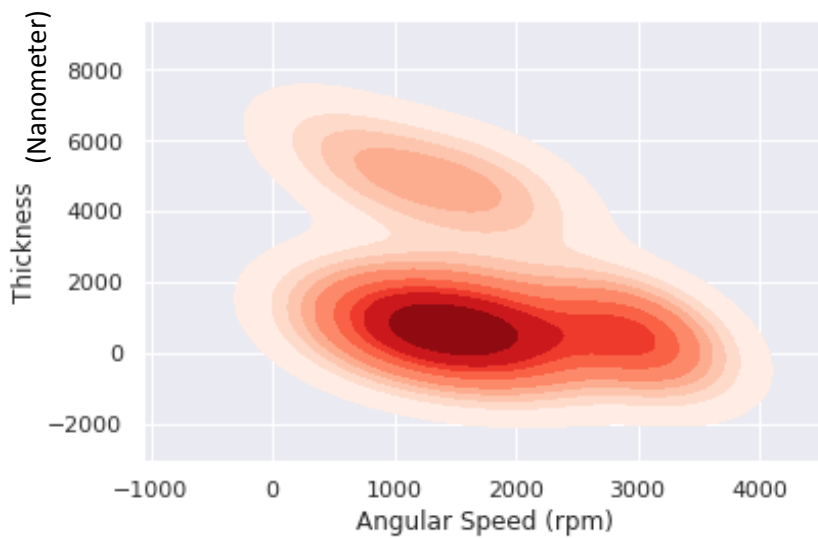

Figure 8. Contour Plot between film thickness and angular speed value

Figure 9 shows the correlation heat map.

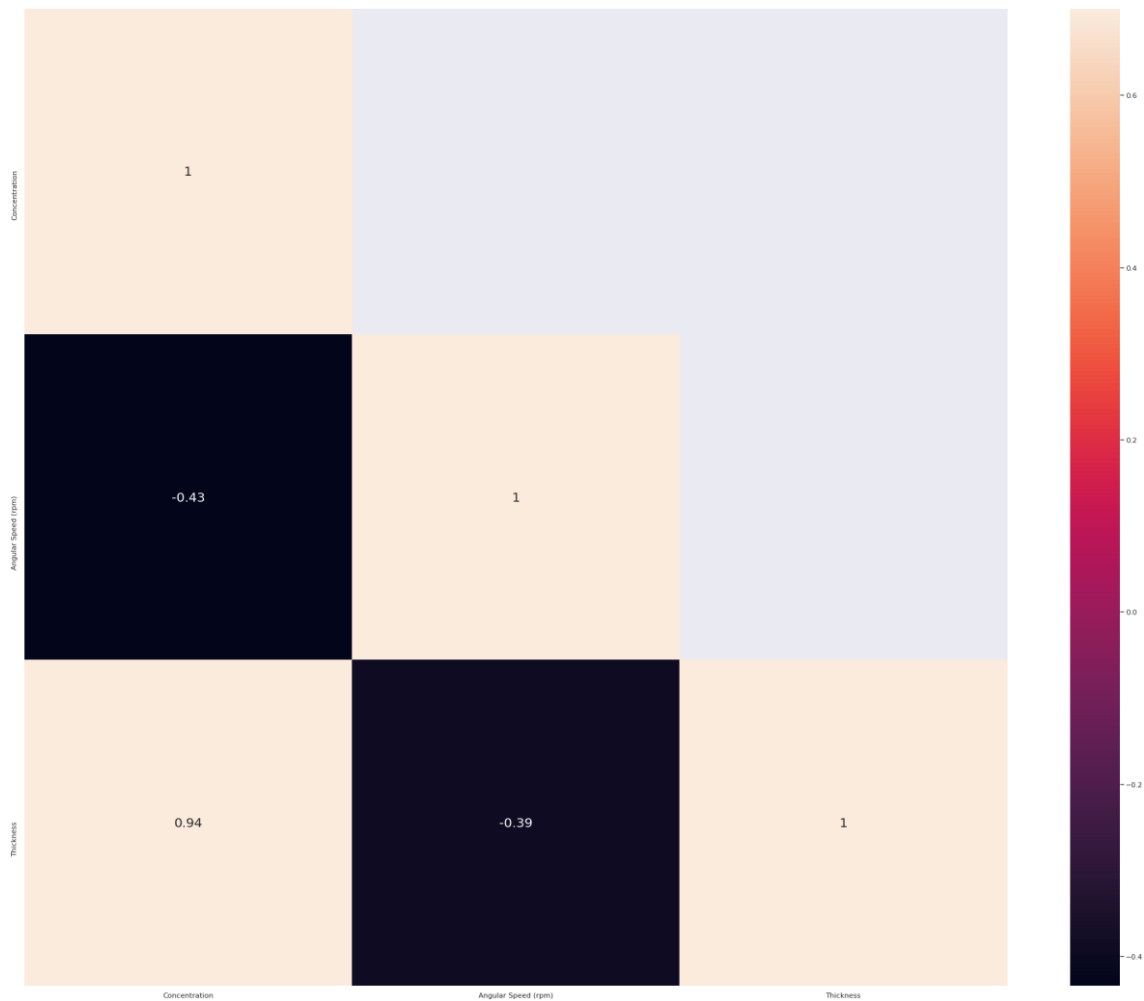

Figure 9. Correlation heat map

3.4 Implementation of Machine Learning Supervised Machine Learning Algorithms

The get_score_regression function will give Mean Absolute Error (MAE), Mean Squared Error (MSE), and R squared error to analyze the model's performance. The Polynomial regression, Decision Tree regression, Random Forest regression, Support Vector Regression and Neural Network regression models were implemented and the one which performs the best was selected. Table 3 shows the model analysis of the regression models implemented on the dataset.

Table 3. Mean Absolute Error, Mean Square Error and Coefficient of determination

|   | Model | MAE | MSE | R2 |
|---|---|---|---|---|
| 0 | Polynomial Regression | 0.150734 | 0.043643 | 0.962665 |
| 1 | SVR | 0.327171 | 0.237411 | 0.796902 |
| 2 | Decision Tree Regressor | 0.122224 | 0.060503 | 0.948242 |
| 3 | Random Forest Regressor | 0.146197 | 0.079901 | 0.931647 |
| 4 | DNN | 0.227174 | 0.214913 | 0.813362 |

Figure 10 shows the model performance of the Neural Network Regressor model. Epoch is defined as the number of times the dataset is passed through the Artificial Neural Network in forward and backward direction.

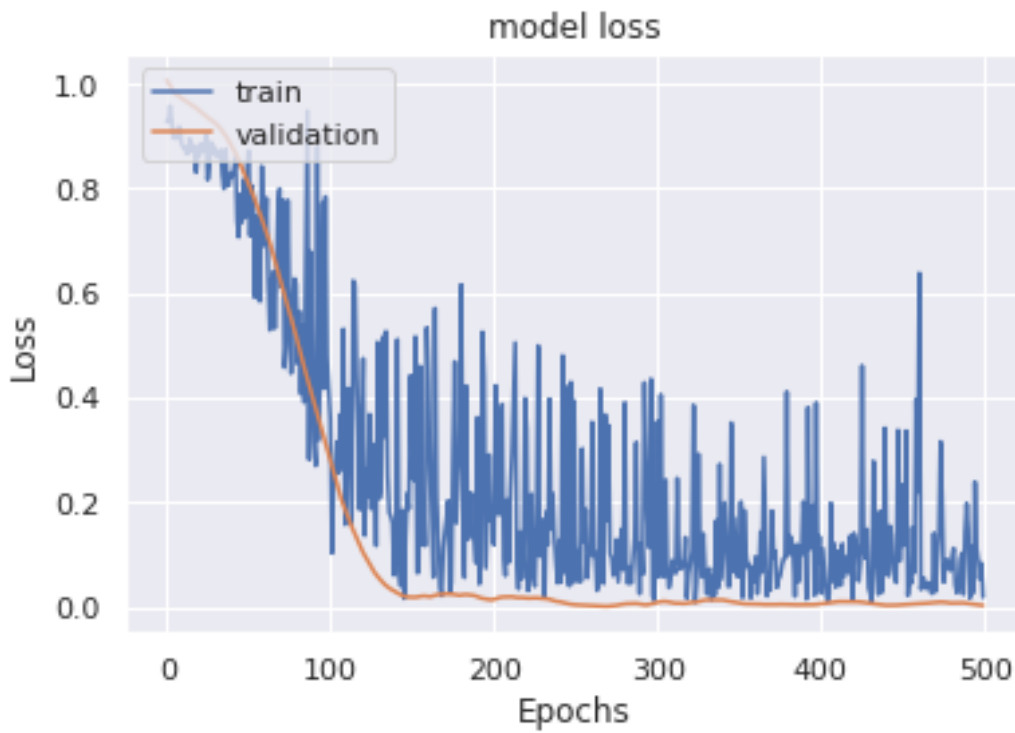

Figure 10. Variation of loss function with number of epochs

Figure 11 and Figure 12 shows the model performance in terms of Mean Square Error (MSE) and Mean Absolute Error (MAE).

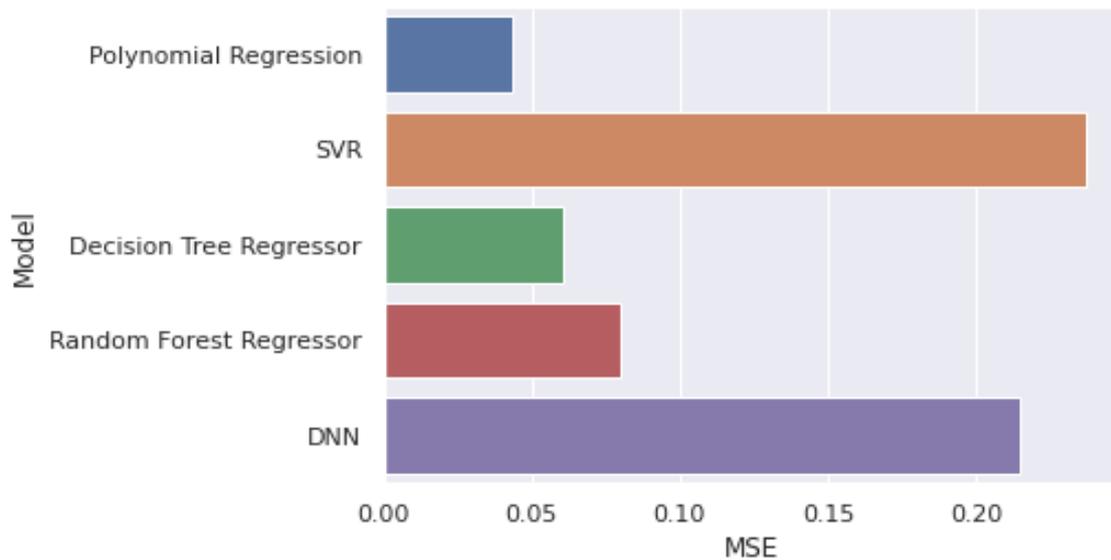

Figure 11. Representation of MSE of each Machine Learning regression models

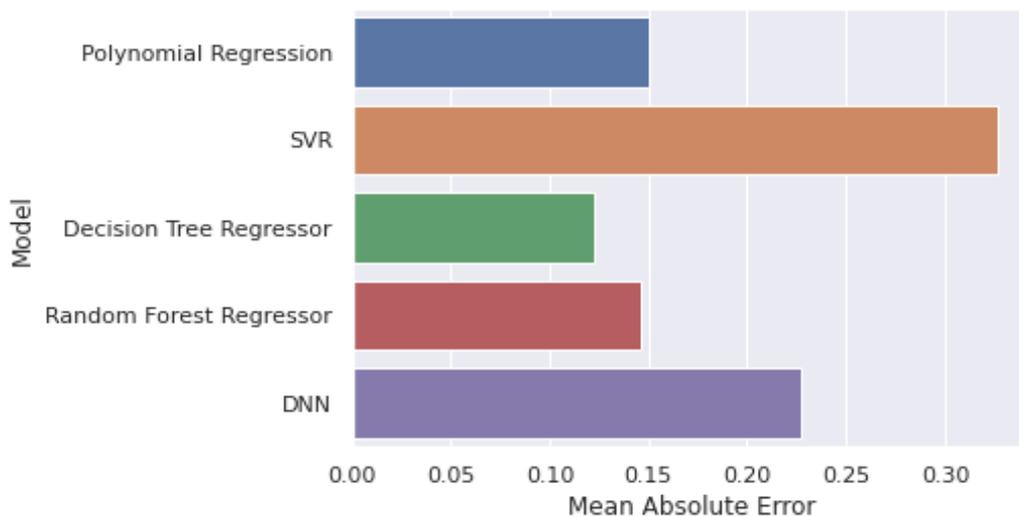

Figure 12. Representation of Mean Absolute Error of each Machine Learning Regression Models

From Figure 11 and 12 it can be clearly interpreted that the Polynomial Regression model and Decision Tree are a better fit than other models while on the basis of Mean absolute error, we can see that the Polynomial Regression model outperforms all other models because of high R square value which is is statistical measurement of how close the data are to the fitted regression line.

## 4. Conclusion

Various supervised machine learning algorithms were successfully implemented on the available thin films dataset. From the obtained results it is observed that the polynomial machine learning model shows the best fit having coefficient of determination of 0.96 approximately while decision tree regression model is the second best model after polynomial regression model having coefficient of determination of 0.94 approximately. It can be concluded that the implemented machine learning algorithms can by simultaneously used for the synthesis of thin films and for thin film design purpose. The implementation of Machine learning algorithms led to reduction in experimental time as well as computational cost. The future work can be based on using more data for the given dataset in order to increase the accuracy of the machine learning algorithms. Also there is a need of implementing quantum machine learning algorithm for further enhancing the accuracy of the obtained results.

**Conflict of Interest**

All authors declare no conflict of interest.